\documentclass[amsmath,amssymb,twocolumn,prc,showpacs,nofootinbib,floatfix,superscriptaddress]{revtex4-1}
\usepackage{graphics}
\usepackage{dcolumn}
\usepackage{bm}
\usepackage{longtable}
\usepackage[dvips]{graphicx}
\usepackage{dcolumn}
\usepackage{footnote}
\usepackage{amssymb}
\usepackage{multirow}
\usepackage{epstopdf}
\usepackage{hyperref}

\begin{document}

\title{Observation of Doppler broadening in $\beta$-delayed proton-$\gamma$ decay}
\author{S.~B.~Schwartz}
\email{schwartz@nscl.msu.edu}
\affiliation{Department of Physics and Astronomy, Michigan State University, East Lansing, Michigan 48824, USA}
\affiliation{National Superconducting Cyclotron Laboratory, Michigan State University, East Lansing, Michigan 48824, USA}
\affiliation{Geology and Physics Department, University of Southern Indiana, Evansville, Indiana 47712, USA}
\author{C.~Wrede}
\email{wrede@nscl.msu.edu}
\affiliation{Department of Physics and Astronomy, Michigan State University, East Lansing, Michigan 48824, USA}
\affiliation{National Superconducting Cyclotron Laboratory, Michigan State University, East Lansing, Michigan 48824, USA}
\author{M.~B.~Bennett}
\affiliation{Department of Physics and Astronomy, Michigan State University, East Lansing, Michigan 48824, USA}
\affiliation{National Superconducting Cyclotron Laboratory, Michigan State University, East Lansing, Michigan 48824, USA}
\author{S.~N.~Liddick}
\affiliation{Department of Chemistry, Michigan State University, East Lansing, Michigan 48824, USA}
\affiliation{National Superconducting Cyclotron Laboratory, Michigan State University, East Lansing, Michigan 48824, USA}
\author{D.~P\'{e}rez-Loureiro}
\affiliation{National Superconducting Cyclotron Laboratory, Michigan State University, East Lansing, Michigan 48824, USA}
\author{A.~Bowe}
\affiliation{Department of Physics and Astronomy, Michigan State University, East Lansing, Michigan 48824, USA}
\affiliation{National Superconducting Cyclotron Laboratory, Michigan State University, East Lansing, Michigan 48824, USA}
\affiliation{Physics Department, Kalamazoo College, Kalamazoo, Michigan 49006, USA}
\author{A.~A.~Chen}
\affiliation{Department of Physics and Astronomy, McMaster University, Hamilton, Ontario L8S 4M1, Canada}
\author{K.~A.~Chipps}
\affiliation{Department of Physics, Colorado School of Mines, Golden, Colorado 08401, USA}
\affiliation{Physics Division, Oak Ridge National Laboratory, Oak Ridge, Tennessee 37831, USA}
\affiliation{Department of Physics and Astronomy, University of Tennessee, Knoxville, Tennessee 37996, USA}
\author{N.~Cooper}
\affiliation{Department of Physics and Wright Nuclear Structure Laboratory, Yale University, New Haven, Connecticut 06520, USA}
\author{D.~Irvine}
\affiliation{Department of Physics and Astronomy, McMaster University, Hamilton, Ontario L8S 4M1, Canada}
\author{E.~McNeice}
\affiliation{Department of Physics and Astronomy, McMaster University, Hamilton, Ontario L8S 4M1, Canada}
\author{F.~Montes}
\affiliation{National Superconducting Cyclotron Laboratory, Michigan State University, East Lansing, Michigan 48824, USA}
\affiliation{Joint Institute for Nuclear Astrophysics, Michigan State University, East Lansing, Michigan 48824, USA}
\author{F.~Naqvi}
\affiliation{Department of Physics and Wright Nuclear Structure Laboratory, Yale University, New Haven, Connecticut 06520, USA}
\author{R.~Ortez}
\affiliation{Department of Physics and Astronomy, Michigan State University, East Lansing, Michigan 48824, USA}
\affiliation{National Superconducting Cyclotron Laboratory, Michigan State University, East Lansing, Michigan 48824, USA}
\affiliation{Department of Physics, University of Washington, Seattle, Washington 98195, USA}
\author{S.~D.~Pain}
\affiliation{Physics Division, Oak Ridge National Laboratory, Oak Ridge, Tennessee 37831, USA}
\author{J.~Pereira}
\affiliation{National Superconducting Cyclotron Laboratory, Michigan State University, East Lansing, Michigan 48824, USA}
\affiliation{Joint Institute for Nuclear Astrophysics, Michigan State University, East Lansing, Michigan 48824, USA}
\author{C.~Prokop}
\affiliation{Department of Chemistry, Michigan State University, East Lansing, Michigan 48824, USA}
\affiliation{National Superconducting Cyclotron Laboratory, Michigan State University, East Lansing, Michigan 48824, USA}
\author{J.~Quaglia}
\affiliation{Department of Electrical Engineering, Michigan State University, East Lansing, Michigan 48824, USA}
\affiliation{Joint Institute for Nuclear Astrophysics, Michigan State University, East Lansing, Michigan 48824, USA}
\affiliation{National Superconducting Cyclotron Laboratory, Michigan State University, East Lansing, Michigan 48824, USA}
\author{S.~J.~Quinn}
\affiliation{Department of Physics and Astronomy, Michigan State University, East Lansing, Michigan 48824, USA}
\affiliation{National Superconducting Cyclotron Laboratory, Michigan State University, East Lansing, Michigan 48824, USA}
\affiliation{Joint Institute for Nuclear Astrophysics, Michigan State University, East Lansing, Michigan 48824, USA}
\author{J.~Sakstrup}
\affiliation{Department of Physics and Astronomy, Michigan State University, East Lansing, Michigan 48824, USA}
\affiliation{National Superconducting Cyclotron Laboratory, Michigan State University, East Lansing, Michigan 48824, USA}
\author{M.~Santia}
\affiliation{Department of Physics and Astronomy, Michigan State University, East Lansing, Michigan 48824, USA}
\affiliation{National Superconducting Cyclotron Laboratory, Michigan State University, East Lansing, Michigan 48824, USA}
\author{S.~Shanab}
\affiliation{Department of Physics and Astronomy, Michigan State University, East Lansing, Michigan 48824, USA}
\affiliation{National Superconducting Cyclotron Laboratory, Michigan State University, East Lansing, Michigan 48824, USA}
\author{A.~Simon}
\affiliation{National Superconducting Cyclotron Laboratory, Michigan State University, East Lansing, Michigan 48824, USA}
\affiliation{Department of Physics and Joint Institute for Nuclear Astrophysics, University of Notre Dame, Notre Dame, Indiana 46556, USA}
\author{A.~Spyrou}
\affiliation{Department of Physics and Astronomy, Michigan State University, East Lansing, Michigan 48824, USA}
\affiliation{National Superconducting Cyclotron Laboratory, Michigan State University, East Lansing, Michigan 48824, USA}
\affiliation{Joint Institute for Nuclear Astrophysics, Michigan State University, East Lansing, Michigan 48824, USA}
\author{E.~Thiagalingam}
\affiliation{Department of Physics and Astronomy, McMaster University, Hamilton, Ontario L8S 4M1, Canada}
\date{\today}

\begin{abstract}
\textbf{Background:} The Doppler broadening of $\gamma$-ray peaks due to nuclear recoil from $\beta$-delayed nucleon emission can be used to measure the energies of the nucleons. This method has never been tested using $\beta$-delayed proton emission or applied to a recoil heavier than $A=10$.
\\
 \textbf{Purpose:} To test and apply this Doppler broadening method using $\gamma$-ray peaks from the $^{26}$P($\beta p\gamma$)$^{25}$Al decay sequence.
\\
 \textbf{Methods:} A fast beam of $^{26}$P was implanted into a planar Ge detector, which was used as a $^{26}$P $\beta$-decay trigger. The SeGA array of high-purity Ge detectors was used to detect $\gamma$ rays from the $^{26}$P($\beta p\gamma$)$^{25}$Al decay sequence. 
 \\
 \textbf{Results:} Radiative Doppler broadening in $\beta$-delayed proton-$\gamma$ decay was observed for the first time. The Doppler broadening analysis method was verified using the 1613 keV $\gamma$-ray line for which the proton energies were previously known. The 1776 keV $\gamma$ ray de-exciting the 2720 keV $^{25}$Al level was observed in $^{26}$P($\beta p\gamma$)$^{25}$Al decay for the first time and used to determine that the center-of-mass energy of the proton emission feeding the 2720-keV level is 5.1 $\pm$ 1.0 (\textit{stat.}) $\pm$ 0.6 (\textit{syst.}) MeV, corresponding to a $^{26}$Si excitation energy of 13.3 $\pm$ 1.0 (\textit{stat.}) $\pm$ 0.6 (\textit{syst.}) MeV for the proton-emitting level. 
 \\
 \textbf{Conclusions:} The Doppler broadening method has been demonstrated to provide practical measurements of the energies for $\beta$-delayed nucleon emissions populating excited states of nuclear recoils at least as heavy as $A=25$.

\end{abstract}


\maketitle

Away from the valley of $\beta$ stability on the chart of nuclides, the nucleon-emission energy thresholds tend to become lower until they cross zero at the drip lines. Concurrently, the $\beta$-decay $Q$ values tend to become larger. Combining these trends creates a large probability for $\beta$-delayed particle emission near the drip lines \cite{pf12rmp}. Considering ongoing advancements in rare-isotope beam production techniques it is, therefore, becoming increasingly important to develop experimental methods to measure $\beta$-delayed particle emissions. On the proton-rich side of the valley of $\beta$ stability it is relatively straightforward to measure these emissions directly because the particles are usually charged \cite{bl08ppnp}. However, on the neutron-rich side the detection of uncharged $\beta$-delayed neutrons is much more challenging \cite{da94pne}. Experimental data on these neutrons are necessary to solve long-standing problems in nuclear structure, nuclear astrophysics, and nuclear energy. Considering the complex decay schemes involved and the modest neutron-energy resolution of existing techniques, the development of complementary experimental tools for the study of $\beta$-delayed neutrons is particularly valuable \cite{ye13prl}.

When $\beta$-delayed particle emissions populate excited states a $\beta$-particle-$\gamma$ decay sequence can occur. The emitted particle causes the residual nucleus to recoil with a velocity that depends on the center-of-mass (c.~m.) energy for the particle emission. This velocity can be measured using the Doppler shift of the $\gamma$ ray providing a measure of the c.~m. decay energy. For an ensemble of isotropic, uncorrelated decays in free space, the Doppler shift is manifested as a uniformly broadened peak in the $\gamma$-ray energy spectrum and the degree of broadening can be used to determine the recoil velocity. If the decay occurs in a medium then the recoiling nucleus may slow down before it emits the $\gamma$ ray; in this case knowledge of the lifetime of the $\gamma$-decaying state and the the stopping power can be used to reconstruct the initial recoil velocity. A detailed formalism for the analysis of this kind of Doppler broadening has been presented in Ref. \cite{fy03nim}.

The Doppler broadening of $\gamma$-ray lines produced by the decays of excited states populated in $\beta$-delayed nucleon emission has only been reported in one very light system: the $\beta$-delayed neutron emission of $^{11}$Li \cite{bo97prc,fy03nim,fy04npa,sa04prc,ma09prc}. The application of this method to higher-mass systems is more challenging because the recoil velocities are smaller. Therefore, it is important to test this technique and improve the precision so it can be applied to nuclides with the highest masses possible \cite{fy03nim}.

In the present work, the use of this Doppler-broadening method in the highest-mass system yet is experimentally demonstrated: the $\beta$-delayed proton decay of $^{26}$P to $^{25}$Al. By using a charged-particle emission for the first time, it was possible to test and verify the Doppler broadening analysis technique for a $^{26}$P($\beta p\gamma$)$^{25}$Al $\gamma$-ray transition associated with two well-known proton energies and a well-known lifetime that could be used as constraints. The technique was then applied to measure an unknown proton energy using a newly-discovered $^{26}$P($\beta p\gamma$)$^{25}$Al $\gamma$-ray line and determine the excitation energy of the proton-emitting $^{26}$Si state.

The $^{26}$P $\beta$-decay experiment has already been described in Ref. \cite{be13prl}, which focused on a $^{26}$Si excited state of astrophysical interest. Briefly, $^{26}$P activity of up to 100 decays per second was produced by the in-flight method at the National Superconducting Cyclotron Laboratory using a 75-pnA, 150-MeV/u $^{36}$Ar primary beam and a $^9$Be production target. The $^{26}$P beam was purified using the A1900 fragment separator \cite{mo03nim} and a radio-frequency fragment separator \cite{ba09nim} before being implanted into a planar germanium double-sided strip detector (GeDSSD) \cite{la13nim}, which was used to detect the signals from charged particles including $\beta$ particles. Beam particles were identified by combining the time of flight from a thin scintillator at the A1900 focal plane to two downstream Si detectors with the energy loss in those Si detectors. The average $^{26}$P-beam purity was 74 \% with 18~\% contamination by $^{24}$Al and small fractions of lighter ions. The GeDSSD was surrounded by the SeGA array \cite{mu01nim} of high-purity germanium detectors, which was used to detect $\gamma$ rays. The NSCL digital data acquisition \cite{st09nim,pr14nim} was employed.

\begin{figure*}
 \includegraphics[width=18cm]{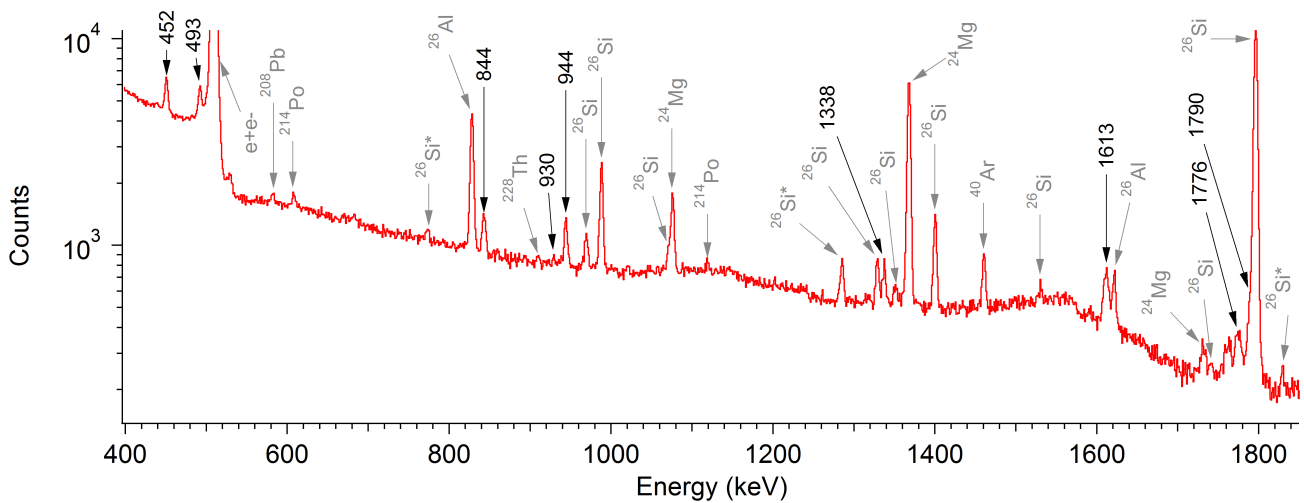}
  \caption{(Color online) $^{26}$P $\beta$-delayed $\gamma$-ray spectrum. All $\gamma$-ray peaks attributed to the $^{26}$P($\beta^{+}p\gamma$)$^{25}$Al decay are labeled by their energy in keV (black). Other peaks are labeled by the $\gamma$-ray emitting nuclide, with escape peaks denoted by an asterisk (gray). Selected regions are shown in more detail in Figs. \ref{fig:1612keV} and \ref{fig:1775keV}.}
   \label{fig:gamma-ray}
\end{figure*}

Signals from the GeDSSD were used to indicate when a $^{26}$P $\beta$ decay took place. The SeGA array was used to detect $\beta$-delayed $\gamma$ rays in coincidence with these signals within a 1.2-$\mu$s timing gate. The SeGA spectra were gain-matched run by run using well-known room-background activity detected in the $\gamma$-ray singles spectrum at 1460.8 keV (from $^{40}$K decay) and 2614.5 keV (from $^{208}$Tl decay). The sum of the gain-matched scintillator-gated SeGA spectra is plotted in Fig. \ref{fig:gamma-ray}.

In the $^{26}$P $\beta$-delayed $\gamma$-ray spectrum (Fig. \ref{fig:gamma-ray}), peaks were observed at 452, 493, 844, 930, 944, 1338, 1613, 1776, and 1790 keV (Table~\ref{tab:intensities}) and attributed to $^{25}$Al $\gamma$ decays produced by the {$^{26}$P($\beta p\gamma$)$^{25}$Al decay sequence. $^{25}$Al peaks were initially identified by searching for the $\gamma$ rays previously observed from this decay \cite{th04epj}. New peaks at 930 and 1776 keV were identified to be candidate $^{25}$Al $\gamma$-ray lines by comparing to the energies of known $^{25}$Al $\gamma$-ray transitions from Ref. \cite{fi09nds}. Coincidences linking several of these $\gamma$ rays in cascades were also observed (Table~\ref{tab:intensities}).

Exponentially modified Gaussian (EMG) response functions were used to fit the peaks of interest. The parameters describing the width of the Gaussian component and the decay of the exponential component were determined as a function of energy by fitting narrow, isolated peaks. Peak centroids and integrals were extracted from the fits, accounting for Doppler broadening when necessary, as described below. 

A linear energy-calibration function was created using well-known $^{24}$Mg $\gamma$-ray energies from $\beta$ decay of the $^{24}$Al beam contaminant \cite{fi07nds}. The calibration was verified to have an accuracy of 0.5 keV using well-known room background peaks in the $\gamma$-ray singles spectrum. The $\gamma$-ray energies are reported in Table \ref{tab:intensities}. The 844-keV peak contains a small contribution from an unresolved $^{26}$P( $\beta \gamma$)$^{26}$Si line at nearly the same energy. In this case, we report the energy of the combined peak. The energies are all consistent with previously reported values when known \cite{fi09nds}.

\begin{table}
\setlength{\tabcolsep}{3pt}
\caption{$^{26}$P($\beta p\gamma$)$^{25}$Al $\gamma$ rays observed in the present work. The measured $\gamma$-ray energies are reported in the first column with their statistical uncertainties only; the global systematic uncertainty is 0.5 keV. An asterisk denotes $\gamma$ rays observed for the first time in $^{26}$P $\beta$ decay. The $\gamma$-ray intensity per $^{26}$P decay is reported in the second column, where the intensity of the 1613-keV line from Ref. \cite{th04epj} was used for normalization. The third column lists $\gamma$ rays observed in coincidence.}
\label{tab:intensities}
\begin{tabular}{c c c}
\hline \hline
  Energy (keV) &  \shortstack{Intensity (\%)} & $\gamma$-ray coincidences \\
  \hline
  451.9(3)         &  2.6(3)       & 493, 844, 930, 1338, 1776 \\
  493.1(4)         &  2.4(3)       & 452, 844, 1776 \\
  843.5(3)         &  0.8(2)       & 452, 493, 944 \\
  930.4(5)$^{*}$   &  0.09(5)      & 452, 944 \\
  944.4(2)         &  1.2(1)       & 844, 930, 1776 \\
  1338.0(2)        &  0.8(1)       & 452 \\
  1613.1(3)        &  2.2(2)       &  \\
  1775.5(3)$^{*}$  &  1.2(1)       & 452, 493, 944 \\
  1790.2(3)        &  0.8(3)       &  \\
\hline \hline
\end{tabular}
\end{table}

The efficiency of SeGA was found by comparing $GEANT4$ Monte Carlo simulations \cite{ag03nim} to data taken offline using an absolutely calibrated $^{154,155}$Eu source and the relative intensities of the $^{24}$Mg lines from online data. Intensities were found by normalizing to the 1613-keV $\gamma$ ray, which is known to have an absolute intensity of 2.2 $\pm $ 0.2 \% \cite{th04epj} based on the proton feeding of the 1613-keV excited state (Table \ref{tab:intensities}). The intensity of the 844-keV peak and its uncertainty were determined by combining the acquired $^{26}$Si data set with sd-shell model calculations \cite{ABrown, DLP}   to predict, and subtract, the small contribution of 0.33 $\pm$ 0.17 \% from the $^{26}$P($\beta\gamma$)$^{26}$Si line.

\begin{center}
\begin{figure}[h]
\includegraphics[width=8.75cm]{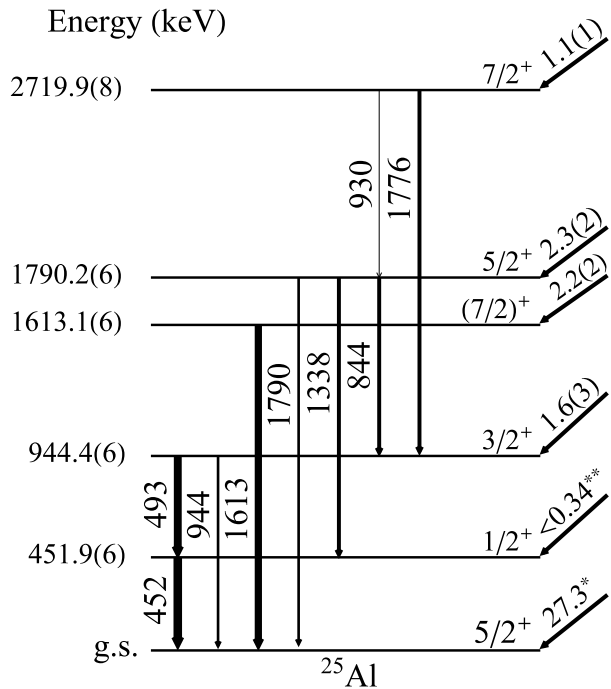}
\caption{$^{25}$Al level scheme from $^{26}$P($\beta p\gamma$)$^{25}$Al decay deduced from the present work. The $\gamma$-ray transitions observed are denoted by arrows with thicknesses proportional to their intensities and labeled by $\gamma$-ray energy in keV. The $\beta p$ feeding of the different excited states is depicted by the arrows on the right, which are labeled by the intensities. The single asterisk denotes a value adopted from Ref. \cite{th04epj}. The double asterisks denotes the upper limit of the $\beta$ feeding of the 452-keV state at the 95\% confidence level.  } 
\label{fig:LevelScheme}
\end{figure}
\end{center}

The $\beta p$ feeding of each $^{25}$Al level was calculated by subtracting the intensity of $\gamma$-decay branches feeding it from the intensity of $\gamma$-decay branches de-exciting it. The $\beta p$ feeding for the ground state has been adopted from previous work \cite{th04epj} since the present experiment was not sensitive to this branch.
 The $\beta$p feeding is summarized in Table \ref{tab:feeding} and illustrated in Fig. \ref{fig:LevelScheme}. We find good agreement with most of the proton-feeding values from Thomas $et$ $al.$, with exception of the feeding of the first excited state at 452 keV. Some of the difference could be attributed to their insensitivity to the 2720-keV state, which provides a significant $\gamma$-ray feeding of the 452-keV state. The small proton feeding can be explained by the need for an $\ell \geq$ 2 proton to populate this $J^{\pi}=1/2^{+}$ state from the $J^{\pi}$=(2,3,4)$^{+}$ $^{26}$Si states fed by $^{26}$P $\beta$ decay.
 

 \begin{center}
 \begin{table}[t]
  \setlength{\tabcolsep}{10pt}
    \caption{The $^{26}$P($\beta p$) feeding of $^{25}$Al states found in the present work and previous work \cite{th04epj}. The 452-keV $\beta$ feeding is given as an upper limit.
    An asterisk denotes evidence for excited states observed for the first time via this decay channel. The intensities are normalized to the feeding of the 1613-keV level from Ref. \cite{th04epj}.}
 \label{tab:feeding}
 \begin{tabular}{c c c}
 \hline\hline

  $^{25}$Al excitation energy (keV)   &   \multicolumn{2}{c}{Proton feeding} \\
                 & \shortstack{Present work\\(\%)} &  \shortstack{ Ref \cite{th04epj}\\(\%)}\\
 \hline
  Ground state   &                 &   27.3 (4)\\
  452            &   $<$0.34         &   2.1  (1)\\
  944            &   1.6(3)        &   2.1  (5)\\
  1613           &   2.2(2)        &   2.2  (2)\\
  1790           &   2.3(2)        &   2.3  (2)\\
  2720*          &   1.1(1)        &           \\
  \hline\hline
 \end{tabular}
 \end{table}
 \end{center}

Doppler broadening was clearly observed in two $\gamma$-ray lines from the $\beta$-delayed proton emission of $^{26}$P to the 1613-keV and 2720-keV excited states of $^{25}$Al. The 1613-keV state de-excites by emitting a 1613-keV $\gamma$ ray (Fig. \ref{fig:1612keV}).  The 2720-keV excited state de-excites predominantly by emitting a 1776-keV $\gamma$ ray (Fig. \ref{fig:1775keV}). Due to the proportionality of the Doppler shift on $\gamma$-ray energy, we were not sensitive to the Doppler broadening of the lines at 1338 keV and below. Due to the proximity of the intense 1797-keV $^{26}$Si peak to the 1790-keV $^{25}$Al peak, it was not possible to study the broadening of the 1790-keV peak precisely.


To analyze the Doppler-broadened peaks we used the method described in Ref. \cite{fy03nim}. A fit function was created that was based on linear combinations of boxcar step functions convoluted with the detector response function. The width of each step function describes the difference between the Doppler shift for $\gamma$-rays emitted by nuclei recoiling directly toward versus away from the detector at a particular speed. Determining the step-function width enabled a calculation of the $^{25}$Al recoil speed at the time the $\gamma$ rays were emitted. The half-life of the $^{25}$Al excited state was incorporated by utilizing the exponential decay equation and the stopping power for Al ions in Ge \cite{zi85srim} to model the deceleration of the recoiling nucleus. The variety of possible recoil speeds introduced by incorporating a finite half-life rounds the shape of the fit function. Other parameters used to describe the broadened $\gamma$-ray lines included the amplitudes and the centroids of the peaks.


The c.~m. energies for the protons that feed the 1613-keV excited state are well known from direct measurements with a silicon detector \cite{th04epj}. This allowed us to test the Doppler broadening method using the 1613-keV peak by adopting these energies and the known lifetime to constrain the fit (Fig. \ref{fig:1612keV}).

The continuous Compton background in the 1613-keV region was modeled with a straight line. The $^{26}$Al peak at 1622.26(3) keV (from $^{26}$Si $\beta$ decay) \cite{en98npa} was modeled to be intrinsically narrow with the amplitude and the centroid as free parameters. The other background peak in the region is at 1611.807(11) keV \cite{en98npa} and is the result of the well-known $\beta$ decay of $^{25}$Al to excited states of $^{25}$Mg. Thanks to the known absolute intensity of this $\gamma$ ray it was possible to fix the intensity of the corresponding peak to be 0.112 times that of the 1613-keV peak, confirming that the overlap of the weak $^{25}$Mg $\gamma$-ray line to the $^{25}$Al $\gamma$-ray line does not contribute significantly to the broadening of the 1613-keV peak.

\begin{figure}[ht]
\includegraphics[width=8.5cm]{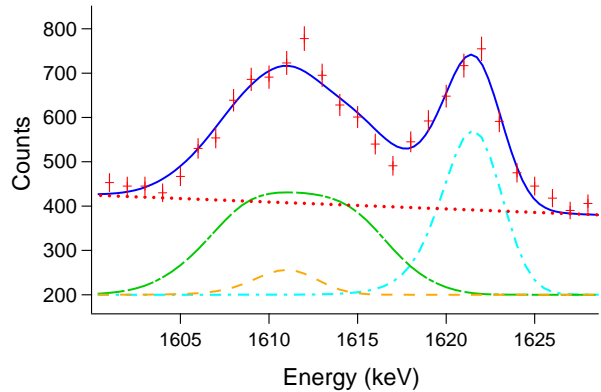}
\caption{(Color online) $^{26}$P $\beta$-delayed $\gamma$-ray spectrum in the region of the $^{25}$Al peak at 1613 keV. The peak at 1613 keV is broader than the neighboring $^{26}$Al peak at 1622 keV. The solid blue (gray) line is the overall fit including Doppler broadening and the dotted red (gray) line represents the Compton scattering background. Below the data and fit, the individual peak components are shown. The 1611-keV $^{25}$Mg, 1613-keV $^{25}$Al, and 1622-keV $^{26}$Al $\gamma$-ray lines are represented by the dot-double-dashed green (gray), dashed gold (gray) and dot-dashed light blue (gray) lines, respectively.}
\label{fig:1612keV}
\end{figure}


The 1613-keV peak was initially assumed to be a narrow line without any Doppler broadening, yielding a $\chi^2$ per degree of freedom of 138/27 corresponding to a $p$ value of 0.0001. This was clear evidence that the 1613-keV peak is not narrow.

A more detailed fit was then performed, which incorporated the Doppler broadening due to the proton emission, including the 12(2)-fs half-life \cite{fi09nds}. There are two different excited states above the proton threshold in $^{26}$Si that emit protons populating the 1613-keV excited state of $^{25}$Al, which undergoes a $\gamma$-ray transition to the ground state \cite{th04epj}. The c.~m. energies of the two protons are 2288(3) keV and 5893(4) keV and they have a relative intensity of $I_{2288}$/$I_{5893}$ = 2.0 (Fig. \ref{fig:decayScheme}) \cite{th04epj}. The fit function was constructed to be a linear combination of these two components. The improvement of the fit after including Doppler broadening with these known values was visually clear (Fig. \ref{fig:1612keV}) and is reflected in the improvement in the $\chi^2$ per degree of freedom to 31.5/27 corresponding to a $p$ value of 0.25. This confirmed the Doppler broadening of this line as well as the accuracy of the Doppler broadening analysis technique \cite{fy03nim} and encouraged a measurement of an unknown proton energy using this method.


\begin{figure}[h]
\includegraphics[width=8.5cm]{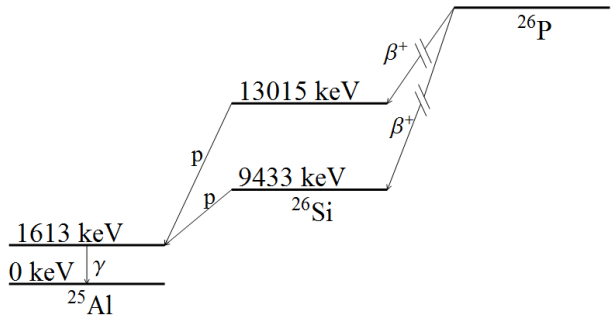}
\caption{The previously known decay scheme for the $^{26}$P $\beta$-delayed proton emission to the 1613-keV excited state of $^{25}$Al \cite{th04epj}. Two proton-unbound $^{26}$Si states feed the 1613-keV $^{25}$Al excited state, causing two different $^{25}$Al recoil velocities following proton emission.}
\label{fig:decayScheme}
\end{figure}

We observed the previously known 1776-keV $\gamma$-ray transition from the 2720-keV excited state of $^{25}$Al for the first time in $^{26}$P $\beta$ decay (Fig. \ref{fig:1775keV}). The c.~m. proton energy feeding the 2720 keV excited state was unknown, enabling the application of the Doppler broadening method to measure the proton energy and identify the proton emitting state of $^{26}$Si.

In the fit of the 1776-keV region, the continuous Compton scattering component of the background was modeled to be linear. Since there were other peaks in the region, it was necessary to include these peaks in the fit function to provide an accurate representation of the background underneath the 1776-keV peak. The 1790-keV $^{25}$Al peak on the shoulder of the strong 1797-keV $^{26}$Si line was modeled to be Doppler broadened; its shape was constrained using the three known proton energies \cite{th04epj} feeding it and the known half-life \cite{fi09nds}.

\begin{figure}
\includegraphics[width=8.5cm]{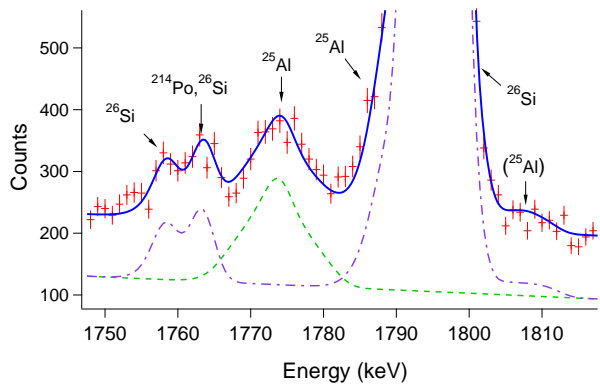}
\caption{(Color online) $^{26}$P $\beta$-delayed $\gamma$-ray spectrum (red [gray] crosses) in the region of the 1776- and 1790-keV peaks with fit (solid blue [gray] line). The dashed green (gray) line shows the contribution of the Doppler-broadened 1776-keV peak to the overall fit. The dot-dashed purple (gray) line shows the contributions of all other peaks. The small excess near 1810 keV could not be confirmed to be from $^{25}$Al.}
\label{fig:1775keV}
\end{figure}

The 1776-keV peak was modeled with Doppler broadening included by incorporating the 201(14)-fs \cite{fi09nds} half-life of the 2720-keV $^{25}$Al state and treating the recoil velocity as an unknown. We found no evidence for multiple proton energies feeding the 2720-keV excited state of $^{25}$Al at our level of sensitivity. The hypothesized initial $^{25}$Al kinetic energy was varied in the fit from 100 to 325 keV in 15-keV steps. The best fit and $\chi^{2}$ value was found for each of these energies. The optimal $^{25}$Al initial kinetic energy was determined to be 195 $^{+41}_{-50}$ (\textit{stat.}) $\pm$ 18 (\textit{syst.}) keV by finding the minimum value of $\chi^{2}$. This energy corresponds to a c.~m. proton energy of 5.1 $\pm$ 1.0 (\textit{stat.}) $\pm$ 0.6 (\textit{syst.}) MeV and a proton-emitting $^{26}$Si level at an excitation energy of 13.3 $\pm$ 1.0 (\textit{stat.}) $\pm$ 0.6 (\textit{syst.}) MeV. Systematic uncertainties for the proton energy were derived from the uncertainties in the shape parameters of the response function, uncertainties in the stopping power, and uncertainties in the background. A summary of the uncertainties can be found in Table \ref{tab:protonUncertainty}, which shows that the stopping power contributes the dominant systematic uncertainty. The only proton unbound state of $^{26}$Si that is consistent with our measured proton energy is the isobaric analog state (IAS) of $^{26}$P at 13.015(4) MeV (the only known excited state above 10.8 MeV), which is also known to be strongly populated in $^{26}$P $\beta$ decay \cite{th04epj}.





\begin{table}
\caption{Sources of uncertainty in the 5.1 $\pm$ 1.0 (\textit{stat.}) $\pm$ 0.6 (\textit{syst.}) MeV $^{26}$P $\beta$-delayed proton c.~m. energy feeding the 2.72-MeV $^{25}$Al state.}
\label{tab:protonUncertainty}
\begin{tabular}{c c}
\hline\hline

\shortstack{Source of \\ uncertainty} & \shortstack{Uncertainty\\(MeV)} \\

\hline
statistics        & 1.0 \\
response function & 0.3 \\
stopping power    & 0.5 \\
background        & 0.1 \\
\hline\hline
\end{tabular}
\end{table}

In conclusion, radiative Doppler broadening in $\beta$-delayed proton-$\gamma$ decay has been observed for the first time. The $^{26}$P($\beta p\gamma$)$^{25}$Al charged-particle emission decay enabled a test of the Doppler broadening analysis technique outlined in Ref. \cite{fy03nim}, which was found to provide a good constrained fit of the broadened 1613-keV $\gamma$-ray line. A new $^{26}$P($\beta p\gamma$) line was discovered at 1776 keV and the Doppler broadening technique was successfully applied to determine the unknown proton energy for this case, yielding the excitation energy of the proton-emitting $^{26}$Si state. To our knowledge, the $A=25$ daughter is the heaviest $\beta$ delayed nucleon emission recoil to which this technique has been applied so far. This confirms that Doppler broadening is a promising complementary tool for the interpretation of $\beta$-delayed particle emission data in general and we anticipate that it will be particularly useful in the context of $\beta$-delayed neutron emission measurements \cite{fy03nim}.

This work was supported by the U.S. National Science Foundation under Grants PHY-1102511, No. PHY 0822648, and No. 1062410, and the U.S. National Nuclear Security Agency under Contract No. DE-NA0000979.

\end{document}